\documentclass{elsart3}
\usepackage{amssymb}
   
\usepackage{graphicx}
\usepackage{dcolumn}
\usepackage{bm}

\begin{document}
 
\begin{frontmatter}

\title{Metal-insulator transition in the In/Si(111) surface}
 
\author[1,2]{S. Riikonen},
\ead{swbriris@sc.ehu.es}
\author[2]{A. Ayuela},
\ead{swxayfea@sc.ehu.es}
\author[3,2]{ D. S\'anchez-Portal}
\ead{sqbsapod@sc.ehu.es}
\address[1]{
Departamento de F\'{\i}sica de Materiales, Facultad
de Qu\'{\i}mica, Universidad del Pa\'{\i}s Vasco, Apdo. 1072, 20080
Donostia-San Sebasti\'an, Spain}
\address[2]{Donostia International Physics Centre (DIPC),
Paseo Manuel de Lardizabal 4,
20018 Donostia-San Sebasti\'an, Spain }
\address[3]{
Centro Mixto CSIC-UPV/EHU, ``Unidad de F\'{\i}sica
de Materiales", Apdo. 1072, 20080 Donostia-San Sebasti\'an, Spain}

\begin{abstract}
The metal-insulator transition observed in the 
In/Si(111)-4$\times$1 reconstruction is studied by means of 
{\it ab initio} calculations of a simplified model of the surface.
Different surface bands are
identified and classified according to their origin
and their response to several structural distortions.
We support the, recently proposed 
[New J. of Phys. 7 (2005) 100], combination
of a shear and a Peierls distortions as the origin of 
the metal-insulator transition.
Our results also seem to favor
an electronic driving force for the transition.
\end{abstract}

\end{frontmatter}
 
Quasi one-dimensional reconstructions formed by metal deposition on 
Si(111) have been intensively studied in recent years. 
The electronic correlations 
and the
coupling between electronic and structural degrees of freedom are
enhanced in one-dimension and, as a consequence, several
electronic and structural phase transitons are observed
in these systems
as the temperature
is decreased. 
A nice example of this behavior is found in the In/Si(111) system,
which exhibits
a 4$\times$1~$\rightarrow$~4$\times$2~$\rightarrow$~8$\times$2
structural transition accompanied by a metal-insulator electronic transition.

The room-temperature (RT) 4$\times$1
structure of the In/Si(111) surface 
is well established~\cite{In_xray,In_coverage}. 
It consists of two neighboring
zigzag In wires along the $[11\bar{2}]$ direction. Each wire
contains two In atoms per 4$\times$1 cell and each In atom
is bonded to one Si atom of the substrate.
This model has been confirmed 
by \emph{ab-initio} calculations \cite{In_dft,In_anis,In_struct} 
which reproduce the scanning tunneling
microscopy (STM) images~\cite{In_stm2,In_stm3}, and 
the main features of the band structure. 

At RT the system presents
three metallic surface bands with similar dispersion~\cite{In_arpes}.
However, when the temperature 
is lowered below $\sim$~130 K~\cite{In_cdw}
photoemission
shows the formation of a band gap. This transition is 
accompanied with a doubling 
of the unit cell in the STM images~\cite{In_cdw}.
The low temperature (LT) phase has 
been widely studied experimentally~\cite{In_low,In_sts,In_coupling,In_opt2}. 
Although several results seem to favor a model where the indium wires suffer a
strong dimerization and break-up into trimers, the question is still far from settled.
In fact, most \emph{ab-initio} calculations find metallic structures 
which are only slightly distorted with respect 
to the RT phase~\cite{In_struct,In_small_gap}. 
Thus, the driving force behind the combined metal-insulator
and structural transition remains unclear.
 
Recently, an interesting mechanism for the gap opening has been proposed 
by Ahn {\it et al.} in Ref.~\cite{In_peierls_gap}. 
The  occupation of the surface
bands in the 4$\times$1 structure is quite close to two electrons. If one
of these bands is depopulated (the upper one),
the other two become very close to half-filled and thus are
suitable to suffer a Peierls transition due to a periodicity doubling. 
If this is true, it seems to indicate that: 
({\it i}) there are at least two types of 
surface bands that originate
in different regions of the substrate or have different symmetries, 
and ({\it ii}) the metal-insulator 
transition is the result of a combination of two distinct structural distortions that
couple with different bands. This last point is consistent 
with the recent first-principles
calculations by Gonz\'alez {\it et. al.}~\cite{In_flores}. 
These authors find an insulating 4$\times$2 structure,
reminiscent of that proposed 
by Kumpf {\it et al.}~\cite{In_xray}, as a result of a combined
shear and Peierls distortion
of the 4$\times$1 RT phase.

The objective of the present work is to understand in detail the origin and
characteristics of the different electronic states involved in the
metal-insulator transition, and how they couple to different
structural distortions. 
The emphasis is on the electronic
bands associated with the indium atoms in the substrate. 
We use a simplified
model (shown in Fig.~\ref{fig:fig1})
that captures the essence of the system.
Our results
support the main conclusions of Ref.~\cite{In_flores}, and point to a 
primary electronic origin of the structural transition.
Of course, in the real surface we can expect
a delicate competition between the gain of
electronic energy and the elastic energy associated 
with the different distortions.
Work along this line is in progress and will be published
elsewhere~\cite{In_our_prb}.

Our calculations were performed with 
the SIESTA 
method~\cite{siesta}.  We used the 
generalized gradient approximation~\cite{gga} 
to the density functional theory, 
norm-conserving pseudopotentials~\cite{tm}, and a 
double-$\zeta$ polarized basis set of numerical atomic orbitals
with cutoff radii corresponding to an 
{\it energy shift} of 200~meV~\cite{siesta}.~\footnote{
The cutoff radius was 4.95~a.u. for the $s$ and $p$ orbitals of H; 
for In, $s$, $p$ and $d$ orbitals had
cutoff radii of 5.95, 7.83, and 7.83 a.u., respectively.}
The density of the auxiliary real-space
grid~\cite{siesta} was equivalent
to a plane-wave cutoff of 100 Ry. We 
typically used 12 inequivalent k-points
along the wire direction (up to 250 in convergence tests, and 
100 in the calculations shown in Fig.~\ref{fig:fig4}). 
We used periodic boundary conditions with a
vacuum distance of 15~\AA.

Figure~\ref{fig:fig1} presents the 
simple model used here to study the electronic properties of 
indium wires in the In/Si(111) surface.
We keep the two zigzag
In wires present in the 4$\times$1 unit cell, and substitute
the neighboring silicon atoms with 
hydrogen. The In-H distances were optimized (1.86 $\AA$), and
the relative positions of the In-H pairs were
kept fixed for the rest of the study.  
This system provides a qualitative model
that, however, retains the main features of the electronic band
structure of the In/Si(111) surface (see below). The In-H bond is 
slightly more ionic than the In-Si bond. However, we have
checked that saturating with SiH$_3$ groups instead of H atoms leads
to a very similar band structure.

Figure~\ref{fig:fig2} shows the evolution of the band structure 
of our model as a function of the wire-wire distance $d$.
There are two types of indium atoms in each wire,
In(1) and In(2). The coordination of the In(2) atoms
changes with $d$. 
Panel (a) corresponds to non-interacting (large $d$) zigzag wires, 
while panel (c) ($d$=2.15~\AA) corresponds to a configuration similar
to that found in the In/Si(111)-4$\times$1 reconstruction.
Although not completely evident due to appearance of interaction gaps,
the band structure in Fig.~\ref{fig:fig2}~(a) 
can be rationalized in terms
of three bands: (I) 
a strongly dispersive band associated 
with the In(1) atoms and the In(2)-p$_y$ orbitals (circles),
a flat band (II) with a clear contribution from 
In(1) and the In(2)-p$_z$ orbitals (triangles), and
another flat band (III) with large 
In(2)-p$_x$ character (squares).
Taken into account the hydrogen saturation, each In atom
contributes with two valence electrons. Thus we have four 
electrons two distribute in these bands. Band (II) is doubly occupied
and does not play any role in the argumentation below. 
Bands (I) and (III), however, are half-filled.
Band (III) can be associated with the 
``dangling-bonds'' in the In(2) atoms that project approximately
into the x-direction and the vacuum.

As $d$ is reduced, the interaction between the wires modifies
the band structure. This can be seen 
in Fig.~\ref{fig:fig2}(b) and (c). 
Particularly, the dispersion of bands derived
from band (III) largely increase
as a result of the overlap of the 
dangling-bonds in the neighboring wires.
Finally, the electronic states associated with 
this band become highly delocalized in the region between 
the two zigzag wires, and the band exhibits an almost
free-electron dispersion. In the following we call
this band the "interaction band".

Figure~\ref{fig:fig2}~(d) presents 
the band structure of the In/Si(111)-4$\times$1
reconstruction calculated with a slab containing 
four silicon bilayers~\cite{In_our_prb}.
Four surface bands can be located in the gap
of the silicon substrate. Three of them cross the 
Fermi level with similar dispersions. 
This is qualitatively reproduced by
our model. The agreement is improved 
if a small shift of the Fermi level
is allowed (see Fig.~\ref{fig:fig2}~(c)).
The information from our model allows to catalogue the
surface bands in two types: ({\it i})
two bands that are derived 
from the band (I) of the right and left wires, 
and ({\it ii}) the interaction band.
The first two bands have a larger
weight {\it inside} each of the wires and 
are quite sensitive to the structure of the zigzag chains.
In contrast, the
interaction band is
localized in the region {\it between} the 
two zigzag wires and thus is more influenced
by the relative positions of the wires.

This division allows to envision a two-step route
for the observed metal-insulator transition.
Step one:
the dimerization of the 
dangling-bonds from neighboring zigzag chains 
opens a gap in the interaction-band. This effect can 
be obtained without  doubling the periodicity of the system, 
the relative displacement of the wires along their axes 
(see Fig.~\ref{fig:fig3}) suffices.
Notice that this corresponds to the shear distortion in Ref.~\cite{In_flores}.
Step two: 
the remaining metallic bands need to
accommodate two valence electrons.
Since these two bands have very similar dispersions, both become
approximately half-filled. As a consequence, the system 
is now suitable to suffer a Peierls transition.
This two-step mechanism is consistent with the
calculations of Ref.~\cite{In_flores} and the experimental 
evidence in Ref.~\cite{In_peierls_gap}.

Figure~\ref{fig:fig3}~(a) shows the evolution of the energy and the 
band structure of our system as function
of the shear distortion.
One of the zigzag wires was displaced 
along the y-direction by a magnitude $\Delta y$ with respect to the other.
For each displacement the distance between the wires $d$ was optimized.
A gap is opened in the interaction band which widens with increasing 
$\Delta y$.
For distortions
larger than $\Delta y\sim$0.5~\AA\ the Fermi level
enters in this gap. This is reflected in the behavior of the
energy that starts to decrease at this point.
This behavior translates in an energy barrier 
of $\sim$5~meV per In atom.
The system is then left
with two metallic bands that cross the Fermi level
at nearby points in reciprocal space.
This is also the case for the real In/Si(111)-4$\times$1 surface 
as can be seen
in Fig.~\ref{fig:fig3}~(b). However, in this case
the behavior of the total energy is different. 
Although still in the range 
of a few meV per In atom for 
moderate distortions, the shear deformation 
always increases the energy of the system~\cite{In_our_prb}.

We now study the effect
of doubling the periodicity along the wires.
We consider a quite simple Peierls-like
distortion:
the length of one every four bonds 
is shorten (the undistorted bond length is 3.045~\AA). 
The distortion is identical for both wires. Still
we have four different possibilities according to the different
relative locations of the distorted bonds
in the neighboring wires.
This is illustrated in Fig.~\ref{fig:fig4}~(a).
While doubling the unit cell, the four distortions 
break the symmetry of the system in different
ways. This detail is quite important. 
The distortions open a gap at ${\pi \over 2a}$ due to the 
periodicity doubling.  However, since the Fermi points do not
exactly lie in that position, this does not guarantee that
the system will become semiconducting.
This is more clear for the extreme shear distortion
($\Delta y$=2.15~\AA, $\alpha$=90$^\circ$).
In this case there is a mirror plane parallel 
to the axis of the wires which 
is only preserved by ``distortion 1''. As a consequence of this 
symmetry, the 
band structure of the system submitted to the ``distortion 1'' presents
a band crossing and the system remains metallic. For the other 
three distortions the band crossing is avoided due to 
the break of symmetry
and a gap opens at the Fermi level. This
can be appreciated in Fig.~\ref{fig:fig4}~(b). 
For $\alpha\neq$90$^\circ$ the symmetry gap is always opened, although
its magnitude depends again on each particular structural distortion.
Figures~\ref{fig:fig4}~(b) and (c) present 
the total energy for the different 
Peierls-like distortions (a 
shear deformation has been previously applied
to the system). ``Distortion 2'' and ``distortion 4'' are always the
most favorables. It is worth noting that the structure proposed
in Ref.~\cite{In_flores} can be understood as the result
of applying a combination of a shear distortion and the Peierls-like
``distortion 2'' presented above.

In summary, we have studied the electronic structure
of the indium zigzag wires seen on the In/Si(111)-4x1
reconstruction. The different surface bands are
identified and classified according to their origin
and their response to different structural distortions. 
We confirm that the combination
of a shear and a Peierls distortion, proposed 
in Ref.~\cite{In_flores},
provides a reasonable and robust route for the
observed metal-insulator transition in this system.
Our results also
point to an electronic driving force of
this transition. 

This work was supported by the Basque Dep. de Educaci\'on and
the UPV/EHU,
the Basque Dep. de Industria (project NANOMAT, ETORTEK
program),
the Spanish MEC, and the
European Network of Excellence FP6-NoE ``NANOQUANTA".
S.R. also acknowledges support 
from the Magnus Ehrnroot foundation.

\bibliographystyle{elsart-num}

\newpage

\begin{figure}\centering
\caption{\label{fig:fig1}Our model of the In wires 
in the In/Si(111)-4$\times$1 surface:
it contains two zigzag indium wires saturated with hydrogen.}
\end{figure}

\begin{figure}\centering
\caption{\label{fig:fig2}
(a)-(c) Band structure (along the wires axes)
of the system shown in Fig.~\ref{fig:fig1} as
a function of the wire-wire distance $d$:
(a) isolated zigzag In wire (d=10.8~\AA),
(b) d=4.31~\AA,  and (c) d=2.15~\AA. Different
symbols indicate the distinct character of the bands
as determined from a Mulliken population 
analysis~\cite{mulliken}:
In(1) and In(2)-p$_y$  
(circles), In(2)-p$_x$ (squares),
and In(1) and In(2)-p$_z$  (triangles). 
(d) Band structure of the In/Si(111)-4$\times$1 
reconstruction along the $[11\bar{2}]$ direction 
calculated with a slab containing
four silicon bilayers~\cite{In_our_prb}, the bands with 
strong indium character are
highlighted with circles. 
Energies are always referred to the Fermi level.}
\end{figure}

\begin{figure}\centering
\caption{\label{fig:fig3}
Shear distortion. (a) Energy per indium atom as a function
of the relative displacement $\Delta y$ of
the indium wires along their axes (see
the scheme in the upper part of panel~(a)).
The band structures for three different values of  $\Delta y$ are also shown.
(b) Band structure of the
In/Si(111)-4$\times$1 reconstruction with $\Delta y$=1.65~\AA.
The inset shows the Brillouin zone, $\Gamma$X and YM
run along the In wires.
Energies in the band structures are referred to the Fermi level. }
\end{figure}

\begin{figure}\centering
\caption{\label{fig:fig4}Peierls distortion.
(a) Scheme showing four inequivalent Peierls-like 
distortions: the 
lenght of one of the bonds (indicated by an arrow) is modified
by the same amount in both wires;
different distortions correspond
to different relative positions of the distorted bonds and 
are numbered according to the labels of the different bonds
in the right wire.
Panels (b) and (c) show the
total energy per indium atom as a function 
of the modified bond length for 
distortions 1 (solid), 2 (dashed), 3 (dotted),
4 (dash-dotted). Panel (b) corresponds to
$\alpha$=90$^\circ$ ($\Delta y$=2.15~\AA),
while in panel (c) $\alpha$=154$^\circ$ ($\Delta y$=1.05~\AA).
The insets show the band structures close 
to the Brillouin-zone boundary 
for distortions 1 and 2 in panel (b), and 2 in panel (c). 
Energies in the band structures are referred to the Fermi level. }
\end{figure}

\end{document}